\documentclass[useAMS,usenatbib]{mn2e}
\usepackage{graphicx,aas_macros,amsbsy,amssymb,amsmath,inputenc,color}%,draftcopy}
\usepackage{multirow}
\usepackage[format=hang,position=bottom,singlelinecheck=false]{caption,subfig}

\title[Density mapping with weak lensing and phase information]{Density mapping with weak lensing and phase information}
\author[R. Szepietowski et al.]{\parbox[t]{\textwidth}{Rafa{\l} M. Szepietowski\thanks{E-mail:
rafal.szepietowski@manchester.ac.uk}$^{1,2}$, David J. Bacon$^1$, J\"org P. Dietrich$^{3,4}$, Michael Busha$^5$, Risa Wechsler$^{6,7,8}$, Peter Melchior$^{9,10}$}\\
\vspace*{4pt} \\
$^1$Institute of Cosmology and Gravitation, University of Portsmouth, Dennis Sciama Building, Burnaby Road, Portsmouth PO1 3FX\\
$^2$Jodrell Bank Centre for Astrophysics, School of Physics and Astronomy, The University of Manchester, Manchester M13 9PL\\
$^3$Universit\"ats-Sternwarte, Ludwig-Maximilians-Universit\"at M\"unchen, Scheinerstr. 1, 81679 M\"unchen, Germany\\
$^4$Excellence Cluster Universe, Boltzmannstr. 2, 85748 Garching b. M\"unchen, Germany\\
$^5$Institute for Theoretical Physics, Universit\"at Z\"urich, 8057 Z\"urich, Switzerland\\
$^6$Kavli Institute for Particle Astrophysics and Cosmology, 452 Lomita Mall, Stanford University, Stanford, CA 94305, USA\\
$^7$Physics Department, Stanford University, Stanford, CA 94305, USA\\
$^8$SLAC National Accelerator Laboratory, 2575 Sand Hill Rd., MS 29, Menlo Park, CA 94025, USA\\
$^9$Center for Cosmology and Astro-Particle Physics, The Ohio State University, 191 W. Woodruff Ave., Columbus, OH 43210, USA\\
$^{10}$Department of Physics, The Ohio State University, 191 W. Woodruff Ave., Columbus, OH 43210, USA}

\date{Accepted 2014 February 25. Received 2014 February 4; in original form 2013 June 21}

\volume{440} \pagerange{2191--2200} \pubyear{2014}

\begin{document}

\label{2191}

\maketitle

\begin{abstract}
The available probes of the large scale structure in the Universe have distinct properties: galaxies are a high resolution but biased tracer of mass, while weak lensing avoids such biases but, due to low signal-to-noise ratio, has poor resolution. We investigate reconstructing the projected density field using the complementarity of weak lensing and galaxy positions. We propose a maximum-probability reconstruction of the 2D lensing convergence with a likelihood term for shear data and a prior on the Fourier phases constructed from the galaxy positions. By considering only the phases of the galaxy field, we evade the unknown value of the bias and allow it to be calibrated by lensing on a mode-by-mode basis. By applying this method to a realistic simulated galaxy shear catalogue, we find that a weak prior on phases provides a good quality reconstruction down to scales beyond $l=1000$, far into the noise domain of the lensing signal alone. 
\end{abstract}


\section{Introduction}

Weak lensing is a promising cosmological probe, allowing the mass distribution in the Universe to be investigated without assumptions about the dynamics of the baryonic component.

In the pioneering work of \citet{kaiser_squires} it has been shown that weak lensing can be used to map the distribution of dark matter in galaxy clusters.  Following this, several methods for making so-called mass maps have been developed, with much attention given to reconstruction methods such as maximum-likelihood approaches \citep{bartelmann}. However, there is a substantial level of noise in the resulting maps, due to the effect of galaxies having intrinsic ellipticities in addition to the sought-after gravitational shear. Therefore it was immediately realised that the reconstruction methods require smoothing or regularisation \citep{squires_kaiser}. A significant proposal in this regard is the Maximum-Entropy method known from image reconstruction studies \citep{bridle,seitz,marshall}. 

These methods work well when applied to clusters, but the lensing ellipticity measurements of galaxies are still sufficiently noisy that reconstruction of the low contrast large-scale structure is not possible with significant signal-to-noise. In this study we develop a methodology attempting to make maps of the projected density with higher signal-to-noise, by utilising a maximum-probability reconstruction with a physically motivated prior probability term: we will examine the usefulness of using Fourier phase information from the distribution of galaxies in the lensing map area. This is related to other recent methods that use galaxy positions to improve density reconstruction \citep{simonb} or combine weak lensing and galaxy positions to measure bias \citep{amara}; in our case, we do not need to assume an amplitude for the bias.

The paper is organised as follows. In Section~2 we review the relevant theoretical background, including weak gravitational lensing quantities and the Fourier description of fields. We also emphasise the importance of Fourier phases in mapping cosmological fields. In Section~3 we introduce the maximum-probability method. We define the likelihood and the prior term for our reconstruction method, describe the phase prior in detail, and outline the practical implementation of our method. Section~4 describes the simulated dataset used in the analysis and the results of applying the reconstruction method. Finally, we discuss the implications of our work in Section~5.

\section{Theory}

\subsection{Lensing quantities}
Here we briefly discuss the necessary lensing theory; full details can be found in e.g. \citet{bartelmann_schneider} and \citet{munshi_heavens}. 

The flat perturbed Friedman-Robertson-Walker metric of the standard cosmological model is 
\begin{equation}
{\rm{d}}s^2=(1+2\Phi/c^2){\rm{d}}t^2-a^2(t)(1-2\Phi/c^2)\big[{\rm{d}}r^2+r^2{\rm{d}}\Omega^2\big],
\end{equation}
where $\Phi$ is the usual Newtonian gravitational potential and $a$ is the scale factor. The potential is related to the matter density field by Poisson's equation
\begin{equation}
\nabla^2_{\rm{com}}\Phi=4\pi G \bar\varrho \delta a^2=\frac{3}{2}H^2_0\Omega_{\rm{m}}\frac{\delta}{a},
\end{equation}
where $\delta=\varrho/\bar\varrho-1$ describes the perturbation around the mean density of matter in the Universe.

In this spacetime a lensing potential can be defined as
\begin{equation}
 \phi(\btheta,r)\equiv \frac{2}{c^2} \int^{r}_{0} {\rm{d}}r' \; \frac{r-r'}{rr'} \; \Phi(\btheta,r'),
\end{equation} 
where $r$ is the comoving distance of the source and the integration is along the line of sight, and $\btheta$ is the position on the sky. This can be understood as a 2-dimensional projection of the gravitational potential. The way in which an image of a source is distorted when passing through a gravitational field depends on a combination of the second order derivatives of the lensing potential
\begin{equation}
 \kappa=\frac{1}{2}(\partial^2_1+\partial^2_2)\phi, 
\end{equation}
\begin{equation}
 \gamma_{1}=\frac{1}{2}(\partial^2_1-\partial^2_2)\phi,
\end{equation}
\begin{equation}
 \gamma_{2}=\partial_1\partial_2\phi,
\end{equation}
where $\kappa$ is called the convergence, $\gamma_1$ and $\gamma_2$ are the two components of the shear $\gamma$, and $\partial_1, \partial_2$ denote angular derivatives in the $x$ and $y$ directions respectively. These quantities are found in the Jacobian matrix 
\begin{equation}
\mathbfss{A}=\left( \begin{array}{cc} {1-\kappa-\gamma_1} & {-\gamma_2} \\ {-\gamma_2} & {1-\kappa+\gamma_1} \end{array} \right),
\end{equation}
which maps the source plane coordinates $\beta_i$ to the image plane coordinates $\theta_j$
\begin{equation}
\mathbfss{A}_{ij}=\frac{\partial\beta_i}{\partial\theta_j}.
\end{equation}
The convergence $\kappa$ describes the projection of the overdensity field on the sky
\begin{equation}
\kappa(\btheta,r)=\frac{3H^2_0\Omega_{\rm{m}}}{2c^2}\int^{r}_{0} {\rm{d}}r'\;\frac{r'(r-r')}{r}\;\frac{\delta(\btheta,r')}{a(r')},
\label{projection}
\end{equation}
and this projected density is the quantity which we seek to reconstruct as a map.

\subsection{Fourier description of fields}
In our reconstruction method, we will use a prior term which involves the phase of lensing fields, so here we define the required quantities for this term. A real space field such as $\kappa$ can be expanded in a Fourier superposition of plane waves:
\begin{equation}
 \kappa(\btheta)=\sum \tilde\kappa(\bmath l)\exp(i \bmath l \cdot \btheta). 
\end{equation}
The Fourier transform $\tilde\kappa$ of such a field is complex and is described by an amplitude $\left|\tilde\kappa(\bmath{l})\right|$ and phase $\alpha_l$ where  
\begin{equation}
 \tilde\kappa(\bmath l)=\left|\tilde\kappa(\bmath l)\right|\exp(i\alpha_l).
\end{equation}
A Gaussian random field will have phases distributed independently\footnote{There is a caveat to this statement. For a real valued field $\kappa$, its Fourier modes have to satisfy the Hermitian relation $\tilde{\kappa}^\ast(l)=\tilde{\kappa}(-l)$.} and uniformly on the interval $[-\pi,\pi)$. The statistical properties of the field are then fully specified by its power spectrum $P(l)=\langle\left|\tilde\kappa(\bmath l)\right|^2\rangle_l$, where $\langle\rangle_l$ denotes an average over all modes at a wavenumber $l$.

However, the phase information contained in the $\kappa$ field is interesting for two reasons:
\begin{itemize}
\item \emph{Morphology}: in cases where one is interested in a specific realisation of a density field, the phases describe features of its spatial pattern \citep{chiang}. For instance, one might be examining a region of the Universe where one wants to know the spatial distribution of matter, to understand the relationship between density and astrophysical properties (e.g. star-formation).
\\
\item \emph{Non-gaussianity}: due to primordial physics \citep[e.g.][]{komatsu} and non-linear evolution on scales probed by weak lensing, the $\kappa$ field will have non-zero higher order statistics beyond the power spectrum. This higher order information is encoded in a combination of phase and amplitude of the Fourier transformed field. If we can obtain a full estimate of phase and amplitude, we will be able to extract information about the growth of structure and the early Universe \citep{watts_coles,chiang_naselsky_coles}.
\end{itemize}

\section{Method}

\subsection{Maximum-Probability reconstruction}

Our reconstruction method seeks to find a hypothesis field which has the maximum probability of accounting for the observed data. We suppose that we have a data vector $\bmath{d}$, which contains estimates of shear from observed galaxy ellipticities. We parameterize the hypothesis field by the values $\bmath{p}$ of projected density in a grid of pixels. The best fitting set of parameters is then found by maximising the posterior probability $P(\bmath{p}|\bmath{d},M)$ according to Bayes' theorem
\begin{equation}
P(\bmath{p}|\bmath{d},M)=\frac{L(\bmath{d}|\bmath{p},M)P(\bmath{p}|M)}{P(\bmath{d}|M)}\propto L(\bmath{d}|\bmath{p},M)P(\bmath{p}|M)
\end{equation}
where $L(\bmath{d}|\bmath{p},M)$ is the likelihood and $P(\bmath{p}|M)$ is the prior probability. The evidence $P(\bmath{d}|M)$ is useful to compare various models $M$, whereas for a particular model $M$ we can simply deal with the proportional term on the right hand side. If we have no knowledge of how the parameters of the model should be distributed, we may assume that all values are equaly likely \emph{a priori} i.e. the prior distribution is flat. Then $P(\bmath{p}|\bmath{d},M)\propto L(\bmath{d}|\bmath{p},M)$ and the posterior distribution is found by maximising the likelihood. This is the basis of maximum-likelihood methods. 

However, the maximum-likelihood method \citep{bartelmann} will typically overfit the data by fitting the noise. Due to finite sampling of the shear field at galaxy positions, and further contamination of the signal by galaxy ellipticity noise, the reconstruction methods require smoothing or regularisation \citep{squires_kaiser}. We can consider two classes of prior which try to achieve this: \emph{informative} and \emph{uninformative} priors, differing in the assumptions which they make about the signal. If the purpose of introducing extra information is to regularise rather than inform an inference we can speak of a \emph{weakly informative} prior. 

Over the past two decades different forms of regularisation have been considered. An important example is the Maximum-Entropy (MaxEnt) regularisation known from image reconstruction \citep{seitz,bridle,marshall} which, while being an uninformative prior, benefitted from inferring information about the correlations in the data \citep{marshall}. In addition, methods have been studied with informative priors; these make some assumptions about the nature of the signal, e.g. Wiener filtering \citep{hu_keeton,simon_taylor_hartlap,simon}. Here we will consider a maximum probability approach with a weakly informative prior.

\subsection{Likelihood}

We would like to find a best fit hypothesized model for the convergence, $\kappa$, given a set of shear observations $\gamma^{\rm{d}}$.  In the flat sky approximation we can relate the convergence and shear fields most easily in Fourier space \citep{kaiser_squires}:
\begin{equation}
\tilde\gamma_1(\bmath{l})=\frac{l_1^2-l_2^2}{l_1^2+l_2^2}\tilde\kappa(\bmath{l}),
\end{equation}
\begin{equation}
\tilde\gamma_2(\bmath{l})=\frac{2l_1l_2}{l_1^2+l_2^2}\tilde\kappa(\bmath{l}).
\end{equation}
As the field of observations will be limited, a simple application of these transformations introduces edge effects, which we will mitigate by making reconstructions over larger patches than the data (see Section~\ref{sec:prct-impl}).

The data vector $\gamma^{\rm{d}}$ consists of estimates of the shear components $\gamma_1$ and $\gamma_2$ in each pixel of a 2D grid. These are obtained by averaging over galaxy ellipticities in each pixel, so that the error on the mean shear in a pixel is 
\begin{equation}
\sigma_{\gamma}\approx\sigma_{\varepsilon}/\sqrt{n}
\label{noise}\end{equation}
where $\sigma_{\varepsilon}$ is the intrinsic  scatter of shear estimators for galaxies, and $n$ the mean number of galaxies in a pixel. This error is approximately Gaussian by the central limit theorem. 

If our hypothesised convergence field has corresponding shear pixel values $\gamma^\kappa_i$, and the data shear pixel values are $\gamma^{\rm{d}}_i$, then the likelihood for our hypothesized reconstruction is:
\begin{equation}
L(\gamma^{\rm{d}}|\kappa)\propto\prod_{i,j}\exp\left(-\frac{(\gamma^{\kappa}_i-\gamma^{\rm{d}}_i)^T \mathbfss{C}^{-1}_{ij}(\gamma^{\kappa}_j-\gamma^{\rm{d}}_j)}{2}\right)
\end{equation}
where $\mathbfss{C}^{-1}$ is the noise covariance matrix. Assuming the noise in each pixel is uncorrelated makes the covariance matrix diagonal and simplifies the likelihood to
\begin{equation}
L(\gamma^{\rm{d}}|\kappa)\propto\prod_i\exp\left(-\frac{(\gamma^{\kappa}_i-\gamma^{\rm{d}}_i)^2}{2\sigma_{\gamma}^2}\right)=\exp\left(-\frac{\chi^2_{\gamma}}{2}\right).
\end{equation}
This assumption is trivially true for shape noise, which dominates on all scales considered. However, intrinsic correlations between galaxy shapes will introduce non-zero off-diagonal terms in the covariance matrix \citep{Catelan01, HirataSeljak04}. 

We turn now to consider the prior term for our maximum-probability reconstruction.

\subsection{Phase prior}
\label{sec:phase_prior}
A prior term that accounts for the claim that galaxies trace mass, even if very poorly, can be achieved by constructing a prediction of the lensing convergence based on galaxy count overdensities
\begin{equation}
 \delta_{\rm{g}}(\btheta,z)=\frac{n_z(\btheta)}{\bar{n}_z}-1,
\end{equation}
where $n_z(\btheta)$ is the number density of galaxies at position $\btheta$ and $\bar{n}_z$ is the mean number density of galaxies at redshift $z$. We could suppose that the overall matter overdensity $\delta\simeq b^{-1}\delta_{\rm{g}}$, where $b$ is the galaxy bias. Then we can project $\delta$ according to Equation \ref{projection} to find the count-estimated convergence $\kappa_{\rm{g}}$. For a sample divided into $N_z$ redshift bins the projection becomes
\begin{equation}
\kappa_{\rm{g}}(\btheta,z)=\frac{3H_{\rm{0}}\Omega_{\rm{m}}}{2c^2}\sum\limits_{i=1}^{N_z}\Delta r_i \frac{r(z_i)[r(z)-r(z_i)]}{r(z)}\frac{\delta(\btheta,z_i)}{a(z_i)},
\end{equation}
where $\Delta r_i=r(z_i)-r(z_{i-1})$. It would then be possible to require that the hypothesized final convergence field is close to this $\kappa_{\rm{g}}$, within some tolerance.

However, there is a problem with this approach: the bias $b$ is unknown, and the claim of linear bias introduces another assumption into the reconstruction.

\begin{figure}
  \centering
    \includegraphics[width=0.5\textwidth]{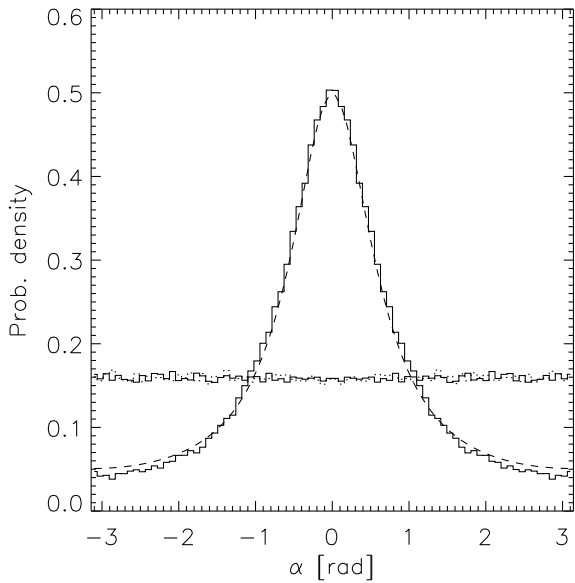}
  \caption{Distribution of convergence Fourier phases and their difference. Histogram of the phases $\alpha$ of the true convergence field $\kappa$ (solid line) and those obtained from the galaxy distribution $\kappa_{\rm{g}}$ (dotted line) for all wavenumbers. The distributions are close to uniform, as expected for fields which have a distribution close to that of a Gaussian random field. Overlaid (peaked curves), the histogram of the phase difference $\Delta\alpha$ between the true convergence $\kappa$ and the approximation $\kappa_{\rm{g}}$ (solid line), for all wavenumbers. The distribution is well approximated by a wrapped Cauchy distribution (dashed line). We see a strong correlation between the phases of the two fields.} 
  \label{ph_distr}
\end{figure}

An easy way of avoiding this problem is to consider only the information about the phases of the Fourier modes of $\kappa_{\rm{g}}$,  neglecting their amplitudes.  Figure \ref{ph_distr} shows the relation between the phases of the true convergence $\kappa$ and count convergence  $\kappa_{\rm{g}}$ found in DES mock catalogue v4.02 (see Section~\ref{sec:data}). 

As expected for a close-to-Gaussian field, the histograms of phases for both $\kappa$ and $\kappa_{\rm{g}}$ fields are close to uniform in the range $[-\pi,\pi)$.  However, the overlaid histogram of the phase difference $\Delta\alpha=\alpha^{\kappa}-\alpha^{\rm{gal}}$ between the true $\kappa$ and $\kappa_{\rm{g}}$ is visibly spiked around $\Delta\alpha=0$, indicating a strong correlation between the phases of the two fields.  We now discuss how this phase difference is calculated in detail.

\begin{figure}
  \centering
    \includegraphics[width=0.5\textwidth]{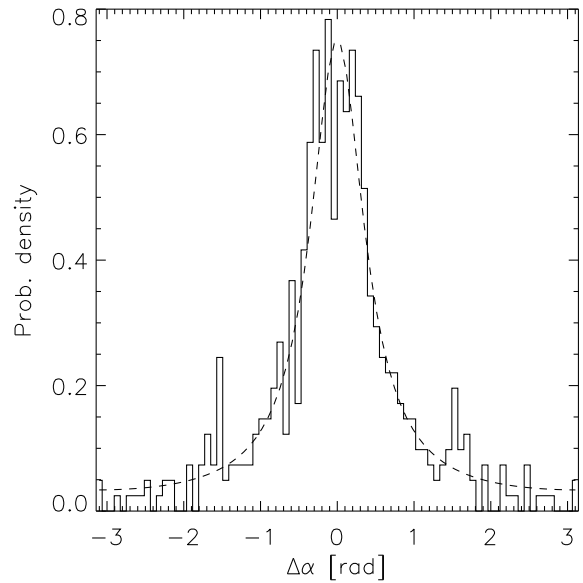}\\
    \includegraphics[width=0.5\textwidth]{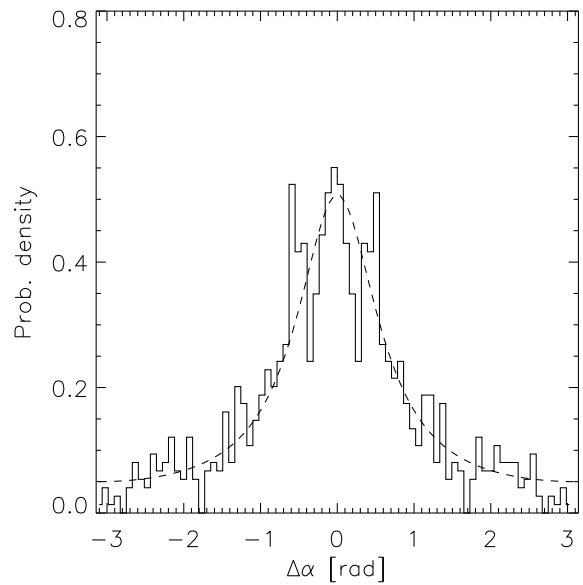}
  \caption{The histogram of the phase difference $\Delta\alpha$ between the true convergence $\kappa$ and the approximation $\kappa_{\rm{g}}$ (solid lines) at $l=1200$ (top) and $l=2250$ (bottom). The distributions are well approximated by a wrapped Cauchy distribution (dashed lines).} 
  \label{ph_l}
\end{figure}

\subsubsection{Phase difference distribution}
\label{sec:phdist}
As the phases are distributed on the interval $[-\pi,\pi)$ their differences will have values on the interval $(-2\pi,2\pi)$. However, since the phases are a cyclic quantity, absolute phase difference $|\Delta\alpha|>\pi$ will correspond to a phase difference smaller than $\pi$. This is easily accounted for: if $\Delta \alpha$ is less than $-\pi$, we add $2\pi$ to $\Delta\alpha$; if $\Delta\alpha$ is greater than or equal to $\pi$ then we subtract $2\pi$ from $\Delta\alpha$.

We can construct the correlation matrix for the phase difference between true convergence phase and galaxy-count derived convergence phase. In our simulations (Section~\ref{sec:data}), this is constructed from 36 different $2^{\circ}\times 2^{\circ}$  areas including $\kappa$ and $\kappa_{\rm{g}}$ information, as for each area only one galaxy distribution realisation is available. By the ergodic principle, this should give an estimate of how much the phases usually differ between the density and galaxy fields in an area. We find that the correlation matrix constructed for $2' \times 2'$ pixels is strongly diagonal with the median absolute value of the correlation coefficient $\simeq 0.06$.

The histograms of $\Delta\alpha$ for the whole field (Figure~\ref{ph_distr}) as well as for individal wavenumbers (Figure~\ref{ph_l}) are well fitted by a wrapped Cauchy probability distribution function:
\begin{equation}
P_{\rm{prior}}(\kappa|\alpha_{\rm{gal}})\propto\prod_i\frac{1-\rho^2}{1+\rho^2-2\rho\cos(\Delta\alpha_i)}.
\end{equation}
We note that the distribution is symmetric around zero. The parameter describing the width of the distribution is $\rho=e^{-\sigma_{\alpha}}$, where $\sigma_{\alpha}$ is the half-width of an unwrapped Cauchy distribution. For small values, $\sigma_{\alpha}$ can be estimated using the median absolute deviation (MAD)
\begin{equation}
\sigma_{\alpha}\approx 1.1\cdot{\rm{MAD}}_{\Delta\alpha}.
\end{equation}
We provide further details on this distribution in Appendix~\ref{wr-c}. However, we want to use the phase information as a weakly informative prior, so we are free to relax this width; we will allow more tolerance in phase difference between our reconstructed $\kappa$ and the $\kappa_g$ field by choosing $\sigma_{\alpha}=2.2\cdot{\rm{MAD}}_{\Delta\alpha}$. Using $\sigma_{\alpha}=1.1\cdot{\rm{MAD}}_{\Delta\alpha}$ would take us in the direction of a \emph{joint} reconstruction of the density field from shear and galaxy position data, which is also of interest; some of our runs in Section~\ref{sec:results} explore this possibility. 

\begin{figure}
  \centering
   \includegraphics[width=0.5\textwidth]{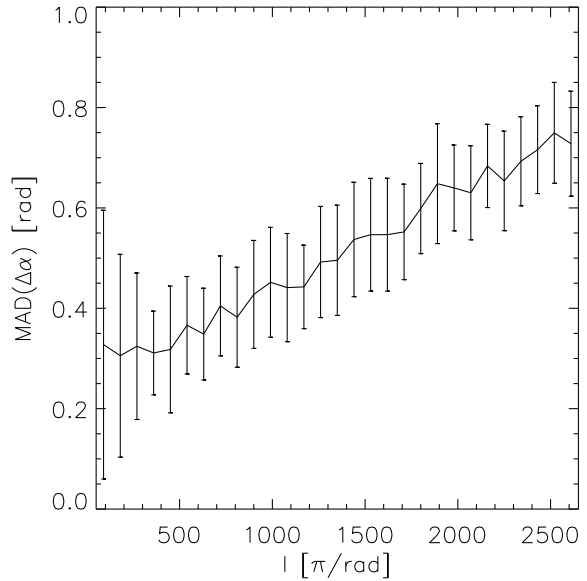}
  \caption{Median absolute deviation (${\rm{MAD}}$) of the phase difference $\Delta\alpha$ between the true convergence $\kappa$ and the approximation obtained from the galaxy distribution $\kappa_{\rm{g}}$ as a function of $l$. The solid line shows the mean ${\rm{MAD}}(\Delta\alpha)$ of the phase difference obtained in shells of radius $l$ from the origin with error bars showing the standard deviation, across the 36 simulated fields.   } 
  \label{ph_error}
\end{figure}

It is to be expected that $\sigma_{\alpha}$ will be a function of $l$, with the phase differences between galaxies and dark matter for large scale modes being more constrained than for small scale ones. We indeed find this to be the case in our simulations, as shown in Figure \ref{ph_error}. The phase difference distribution for each $l$ also follows a wrapped Cauchy distribution. This  distribution is naturally generated when the difference between $\kappa$ and $\kappa_g$ comes from a white noise contribution, such as shot-noise, and possibly a further contribution from the stochasticity of the bias relation \citep{dekel_lahav,manera}. Hence, the low $l$ modes have smaller phase differences, as this white noise offset is smaller as a proportion of the signal on these scales.

In the mock catalogue the galaxy biasing is roughly linear and deterministic. It could be that the wrapped Cauchy pdf of the phase differences is typical only for this type of bias, but might be quite different for more complex scenarios. Hence, further studie of how the phase difference distribution arises are important. However, as we permit very large errors on the phase difference, moderate deviations from our simulations' bias model should not change the conclusions of the paper.

\begin{figure}
  \centering
    \includegraphics[width=0.5\textwidth]{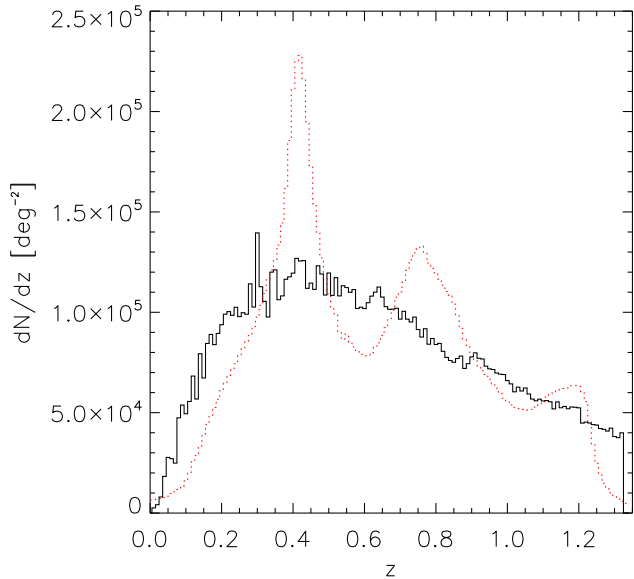}
  \caption{Distribution of galaxies with redshift. True redshift (black solid) and photometric redshifts obtained using the ANN$z$ code (red dotted).} 
  \label{z_distr}
\end{figure}

In reality, the estimation of $\delta_{\rm{g}}$ will suffer from systematics originating, for example, from an inhomogeneous galaxy survey. These could be mitigated by methods used for the matter power spectrum estimation, where pixels are reweighted to account for the mask \citep{FeldmanKaiserPeacock94, PercivalVerdePeacock04}. A further systematic will arise from using photometric redshifts to estimate distances (Figure~\ref{z_distr}). However, this will be mitigated by the fact the convergence is projected; nevertheless, careful tests of this systematic will be necessary.

\subsection{Practical implementation}
\label{sec:prct-impl}
We are now ready to discuss our approach to finding a reconstructed convergence field. Rather than estimating the posterior distribution of our convergence hypotheses, we will seek a \emph{maximum a posteriori} (MAP) solution. The reconstruction is performed by seeking a $\tilde\kappa^{\rm{trial}}$ that maximises the posterior probability. The posterior pdf will be generally strongly peaked so it is convenient to work with its logarithm
\begin{equation}
-\ln P(\tilde\kappa|\gamma^{\rm{d}},\alpha_{\rm{gal}})\propto-\ln L-\ln P_{\alpha}^{\rm{prior}},
\end{equation}
which varies more slowly with the change in $\tilde\kappa$. 

As the shape of the posterior pdf is generally unknown, we use a simple heuristic optimiser. We use the idea of Simulated Annealing \citep{sa}, but replace the usual Metropolis-Hastings sampler \citep{metropolis,hastings} with a Multi Try Metropolis \citep{mtm} one. 
In each step $t$ a set of trial convergence fields $\{\tilde\kappa^{\rm{trial}}_i\}$ is generated from the current field
\begin{equation}
\tilde{\kappa}^{\rm{trial}}_i=\tilde{\kappa}^{\rm{current}}+\delta\tilde{\kappa}_i,
\end{equation}
where components of each $\delta\tilde{\kappa}_i$ are drawn from normal distribution $\mathcal{N}(0,\sigma_t \sqrt{{\rm{P}}(l)})$, where the scaling ${\rm{P}}(l)$ is proportional to the expected signal (see below). A proposal field $\tilde{\kappa}^{\rm{proposal}}$ is then chosen.
To limit the random walk behaviour, the field with the highest probability different from the current one is chosen. Then a reference set $\{\kappa^{\rm{ref}}_j\}$ that includes $\tilde{\kappa}^{\rm{current}}$ is formed from that field. The proposal field is then accepted with the probability
\begin{align}
P(\tilde{\kappa}^{\rm{proposal}}|\{\kappa^{\rm{ref}}_j\})&=1 \quad {\rm{for}} \quad \frac{\sum_j P(\tilde\kappa^{\rm{ref}}_j)}{\sum_i P(\tilde\kappa^{\rm{trial}}_i)} \geq 1, \\
P(\tilde{\kappa}^{\rm{proposal}}|\{\kappa^{\rm{ref}}_j\})&=T_t\frac{\sum_j P(\tilde\kappa^{\rm{ref}}_j)}{\sum_i P(\tilde\kappa^{\rm{trial}}_i)} \quad {\rm{otherwise.}}
\end{align}
In addition to a cooling schedule for the acceptance rate
\begin{equation}
T_t=\frac{T_0}{\log_{10} (t+10)},
\end{equation}
 we have added a similar schedule to decrease the step size in the sampling algorithm
\begin{equation}
\sigma_t=\frac{\sigma_0}{\log_{10} (t^2+10)},
\end{equation}
to allow for more refined changes as the optimiser gets closer to the solution we seek \citep{hybrid, Kotze09}. The solution with the highest probability $\kappa_{\rm{best}}$ is stored and used as the output of the optimiser.

Operations on the fields, such as calculating the shears from the convergence, are performed in Fourier space, hence edge effects such as periodic boundaries of the reconstruction will be present. This would mean that the largest scales would not be recovered accurately. This is partially solved by introducing a larger reconstruction grid as suggested in \citet{bridle} and here we use a grid 4 times bigger than the reconstruction area.

To aid the optimisation process we choose a starting position for our hypothesis which is expected to be close to the MAP solution. The initial guess for the reconstruction, $\tilde\kappa^{\rm{initial}}$, is a field fully consistent with the prior; that is, we choose phases from the galaxy convergence map. We also apply a power spectrum filter to the $\tilde\kappa_{\rm{g}}$ field
\begin{equation}
\tilde\kappa^{\rm{initial}}(\bmath{l})=\tilde{\kappa}_{\rm{g}}(\bmath{l})\sqrt{\frac{{\rm{P}}(l)}{{\rm{P}}_{\rm{g}}(l)}},
\label{filter}
\end{equation}
which gives the $\kappa_{\rm{g}}$ field the required amplitude of power spectrum and suppresses the high-$l$ noise. As this is only a starting guess, any ${\rm{P}}(l)$ with a very approximately correct shape and amplitude should suffice. Here, we choose the true average $\kappa$ power spectrum from simulations. By choosing this starting point, the optimizer evolves the reconstruction from the prior to the posterior under the influence of lensing. 

However, to check for possible local maxima in the posterior, we also try running the code from a noisy position such as $\kappa_{\rm{g}}$ without applying any filters.

\section{Application to simulated data}

\subsection{Simulated galaxy catalogue}
\label{sec:data}

For this study we have used the mock galaxy catalogues created for the Dark Energy Survey based on the algorithm Adding Density Determined GAlaxies to Lightcone Simulations (ADDGALS; Wechsler et al 2013, in preparation; Busha et al 2013, in preparation).  This algorithm attaches synthetic galaxies, including multiband photometry, to dark matter particles in a lightcone output from a dark matter $N$-body simulation and is designed to match the luminosities, colors, and clustering properties of galaxies. The catalogue used here was based on a single ``Carmen'' simulation run as part of the LasDamas of simulations (McBride et al, in preparation)\footnote{Further details regarding the simulations can be found at {\tt http://lss.phy.vanderbilt.edu/lasdamas/simulations.html}}. This simulation modeled a flat $\Lambda$CDM universe with $\Omega_{\rm{m}} = 0.25$ and $\sigma_8 = 0.8$ in a 1 Gpc/$h$ box with $1120^3$ particles. A 220 sq deg light cone extending out to $z = 1.33$ was created by pasting together 40 snapshot outputs.

The galaxy distribution for this mock catalogue was created by first using an input luminosity function to generate a list of galaxies, and then adding the galaxies to the dark matter simulation using an empirically measured relationship between a galaxy's magnitude, redshift, and local dark matter density, $P(M_r,z|\delta_{\rm{dm}})$ -- the probability that a galaxy with magnitude $M_r$ and redshift $z$ resides in a region with local density $\delta_{\rm{dm}}$.  This relation was tuned using a high resolution simulation combined with the SubHalo Abundance Matching technique that has been shown to reproduce the observed galaxy 2-point function to high accuracy \citep{Kravtsov04, Conroy06, Reddick12}.

For the galaxy assignment algorithm, we choose a luminosity function that is similar to the SDSS luminosity function as measured in \cite{Blanton03}, but evolves in such a way as to reproduce the higher redshift observations (e.g., SDSS-Stripe 82, AGES, GAMA, NDWFS and DEEP2). In particular, $\phi_*$ and $M_∗$ are varied as a function of redshift in accordance with the recent results from GAMA \citep{Loveday12}.

Once the galaxy positions have been assigned, photometric properties are added.  Here, we use a training set of spectroscopic galaxies taken from SDSS DR5. For each galaxy in both the training set and simulation we measure $\Delta_5$, the distance to the 5th nearest galaxy on the sky in a redshift bin. Each simulated galaxy is then assigned an SED based on drawing a random training-set galaxy with the appropriate magnitude and local density, k-correcting to the appropriate redshift, and projecting onto the desired filters. When doing the color assignment, the likelihood of assigning a red or a blue galaxy is smoothly varied as a function of redshift in order simultaneously reproduce the observed red fraction at low and high redshifts as observed in SDSS and DEEP2.

For the simulation of gravitational lensing, weak lensing shear at each galaxy position was computed using the multiple plane ray tracing code CALCLENS \citep{becker12}. Then an intrinsic ellipticity is assigned to each galaxy. The intrinsic shape distribution and dispersion $\sigma_\varepsilon$ in these simulations are magnitude dependent and are modeled after those found in deep SuprimeCam i$^\prime$-band data with excellent seeing ($0\farcs6$), with fainter galaxies having a higher intrinsic ellipticity dispersion. Averaged over all galaxies $\sigma_\varepsilon = 0.4$.

\begin{figure*}
  \centering
  \subfloat[True convergence in the simulation.]{
    \includegraphics[width=0.5\textwidth]{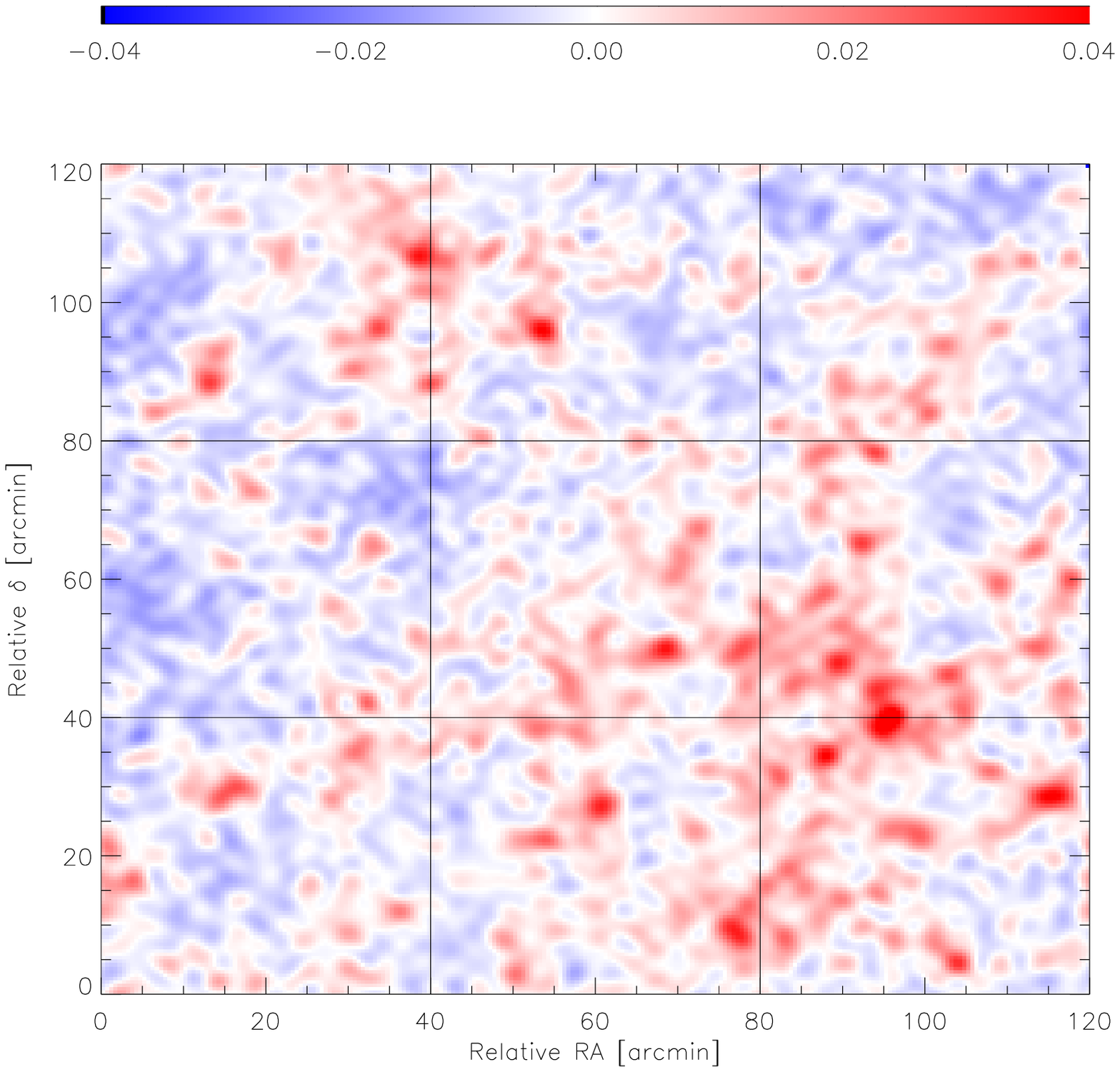}
  } 
  \subfloat[Maximum-likelihood reconstruction.]{
    \includegraphics[width=0.5\textwidth]{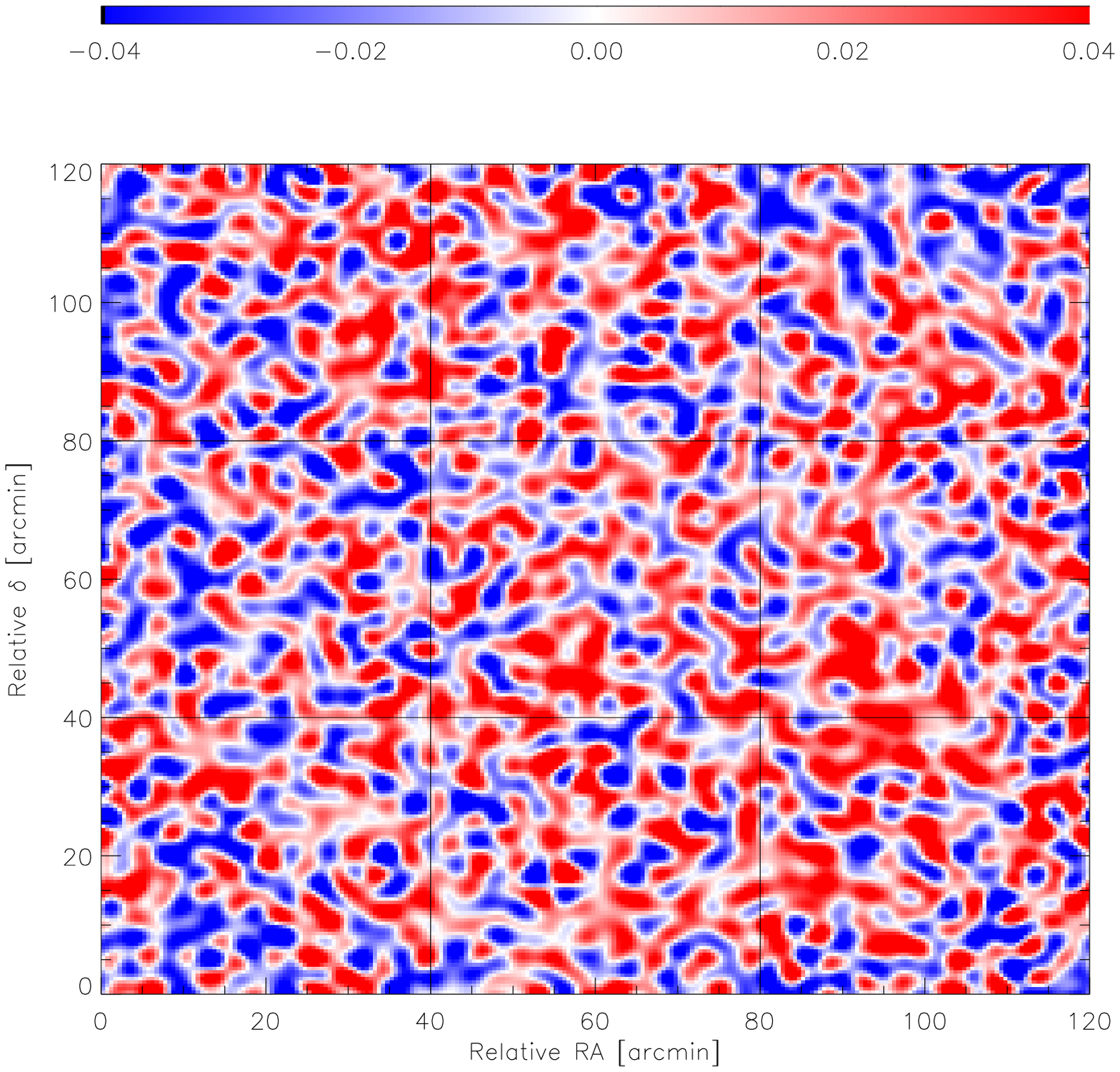}
  }\\ 
\subfloat[Maximum-probability, including phase information.]{
    \includegraphics[width=0.5\textwidth]{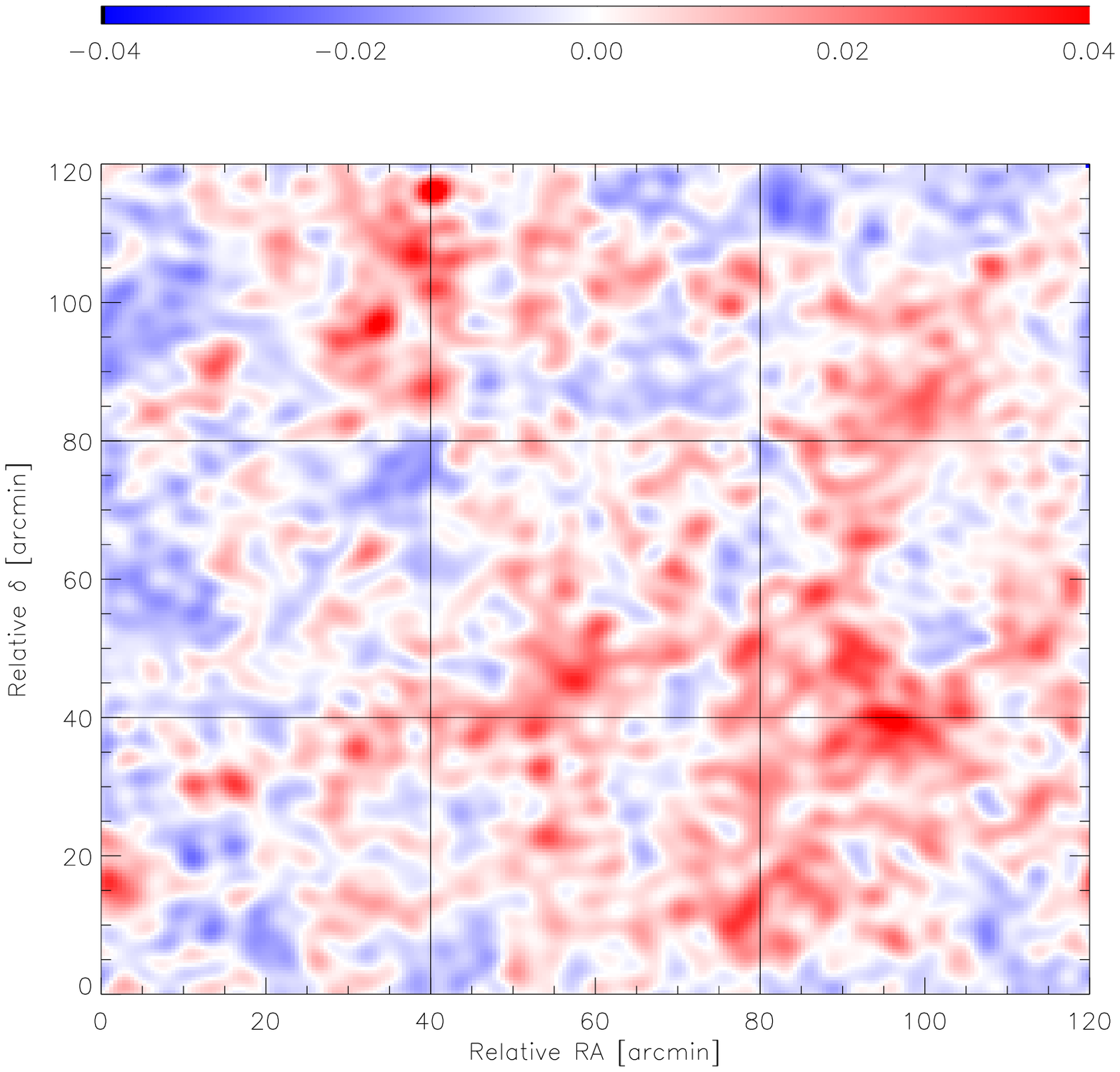}
  }
\subfloat[Convergence estimate from galaxy positions.]{
    \includegraphics[width=0.5\textwidth]{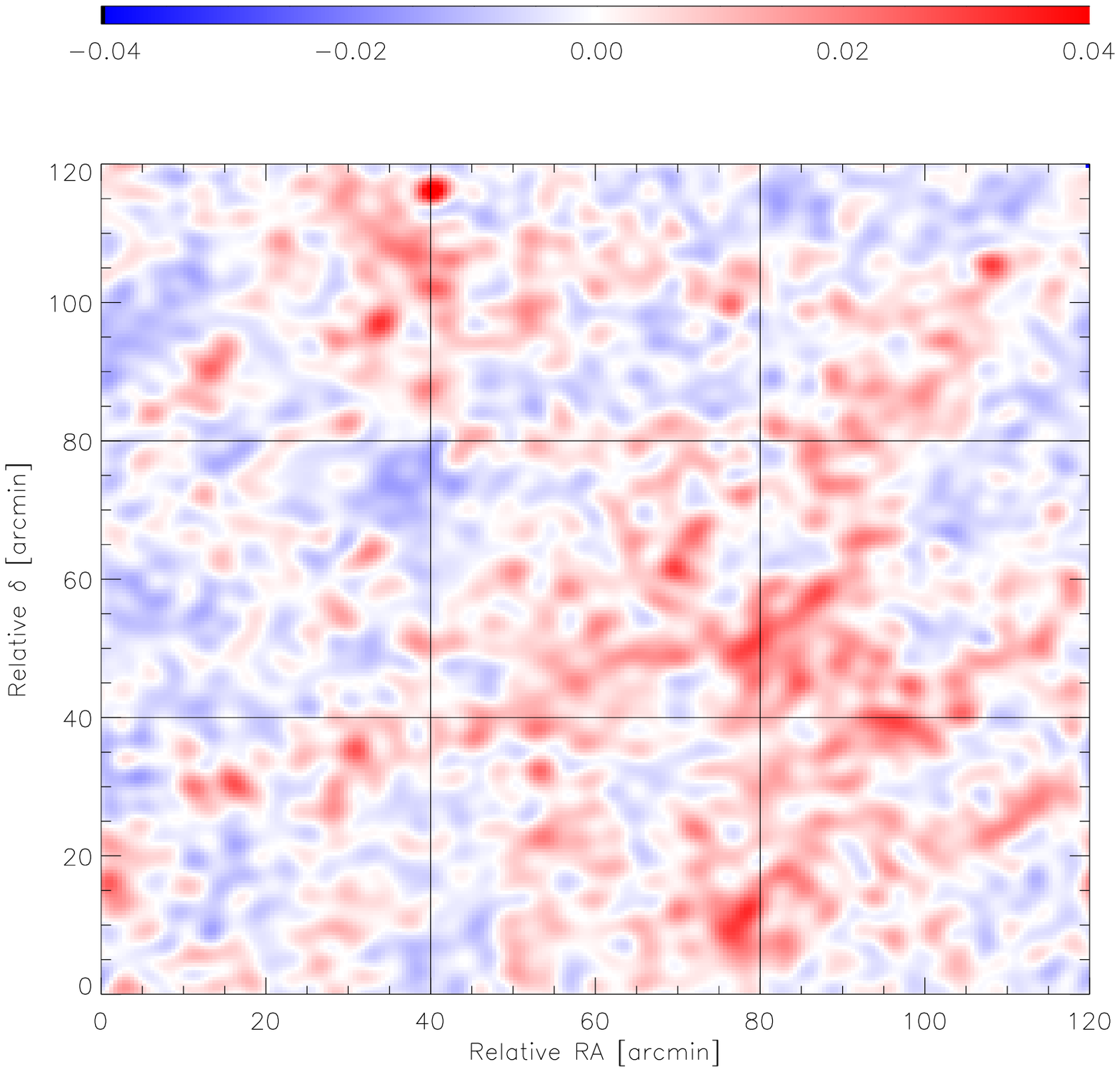}
  }
  \vspace{1cm}
  \caption{Resulting reconstructed maps of the convergence field. The maps are showing an example of a $2^{\circ} \times 2^{\circ}$ reconstruction field with $2^{\prime}\times 2^{\prime}$ pixels. The maps were zero-padded in Fourier space to have a smoother apperance. The true convergence is shown along with reconstructions obtained using the maximum-likelihood method and the maximum-probability method with the phase prior. The galaxy convergence $\kappa_{\rm{g}}$ from which the prior was computed is also shown for comparison.}
  \label{maps}
\end{figure*}

\subsection{Results}
\label{sec:results}

\begin{table}
\centering
\caption{List of the reconstructions carried out, with different combinations of priors, phase distribution parameters and initial reconstruction hypothesis.}
\begin{tabular}{c c c}
\hline \hline
Posterior & Phases tolerance & $\kappa^{\rm{initial}}$ \\ \hline 
$L(\gamma^{\rm{d}}|\kappa)$ & $-$ & Filt. \\\\
$L(\gamma^{\rm{d}}|\kappa)P(\kappa|\alpha_{\rm{gal}})$ & $\sigma_\alpha=1.1\cdot{\rm{MAD}}_{\Delta\alpha}$ & Filt. \\\\
$L(\gamma^{\rm{d}}|\kappa)P(\kappa|\alpha_{\rm{gal}})$ & $\sigma_\alpha=2.2\cdot{\rm{MAD}}_{\Delta\alpha}$ & Filt. \\\\
$L(\gamma^{\rm{d}}|\kappa)P(\kappa|\alpha_{\rm{gal}})$ & $\sigma_\alpha=2.2\cdot{\rm{MAD}}_{\Delta\alpha}$ & Noisy \\\\
$L(\gamma^{\rm{d}}|\kappa)P(\kappa|\alpha_{\rm{gal}})$ & $\sigma_\alpha=3.3\cdot{\rm{MAD}}_{\Delta\alpha}$ & Filt. \\
\hline
\end{tabular}
\label{runs}
\end{table}

From the simulated catalogue described in Section~\ref{sec:data}, we select a large square square patch of $12^{\circ} \times 12^{\circ}$. To study the behaviour of the reconstructions, $100$ areas (with replacement) of $2^{\circ} \times 2^{\circ}$ were randomly selected from this patch. These were divided into pixels of $2^{\prime}\times 2^{\prime}$ containing $\simeq116$ galaxies. Hence the number density of sources is $29\;{\rm{gal}}/{\rm{arcmin}}^2$. We use the same galaxies as sources and tracers of the density field.

The reconstruction code was run for 30,000 trial steps for each sub-field, with $300$ trial fields generated in each optimization step. The reconstructed maps span $4^{\circ} \times 4^{\circ}$, containing 14,400 pixels of $2^{\prime}\times 2^{\prime}$; i.e. we reconstruct a larger patch than the $2^{\circ} \times 2^{\circ}$ data patch in each case.

The reconstructions were performed for each of the $100$ fields using different phase distribution parameters and initial guesses that are summarised in Table \ref{runs}. Using 100 different fields allowed us to examine the noise properties of the reconstruction method.

Reconstructions were performed using a maximum-likelihood (ML) method (i.e. no prior terms) and the maximum-probability approach with the phase prior. In this set of runs, the phase prior included a phase tolerance $\sigma_\alpha=2.2\cdot{\rm{MAD}}_{\Delta\alpha}$ in order to provide a weakly informative prior. To obtain a reasonable starting point, $\tilde\kappa^{\rm{initial}}$ was filtered according to equation (\ref{filter}).

Figure \ref{maps} shows examples of maps obtained using both methods of reconstruction (b and c) with the true simulated convergence map (a) and the convergence estimated from galaxy positions (d) shown for comparison (using $\delta=\delta_{\rm{g}}$, i.e. $b=1$, see Section~\ref{sec:phase_prior}). The ML method reconstructs only the most prominent peaks, with a high level of contamination by spurious peaks. The inclusion of the phases prior appears to improve the map considerably, but it also maps features from $\kappa_{\rm{g}}$ that are not necessarily present in the true convergence, e.g. ${\rm{RA}}=40^{\prime}, \; \delta=115^{\prime}$. However, these are consistent with the lensing only reconstruction.

To quantify the quality of the reconstruction, we construct a power spectrum of the error per mode in the reconstruction,
\begin{equation}
{\rm{P}}_{\rm err}(l) = \langle | \tilde\kappa^{\rm{rec}}_l -\tilde\kappa^{\rm true}_l  |^2 \rangle_l.
\end{equation}
A faithful reconstruction will have small ${\rm{P}}_{\rm err}(l)$, preferably smaller than the true power in order to achieve good $S/N$ (i.e. the errors in the reconstruction are preferably smaller than the signal of the reconstructed structures for a given scale). ${\rm{P}}_{\rm err}(l)$ shows the scale dependence of the reconstruction faithfulness. However, it is not intended as a metric of how well we can reconstruct the power spectrum from the maps.

\begin{figure}
  \includegraphics[width=0.5\textwidth]{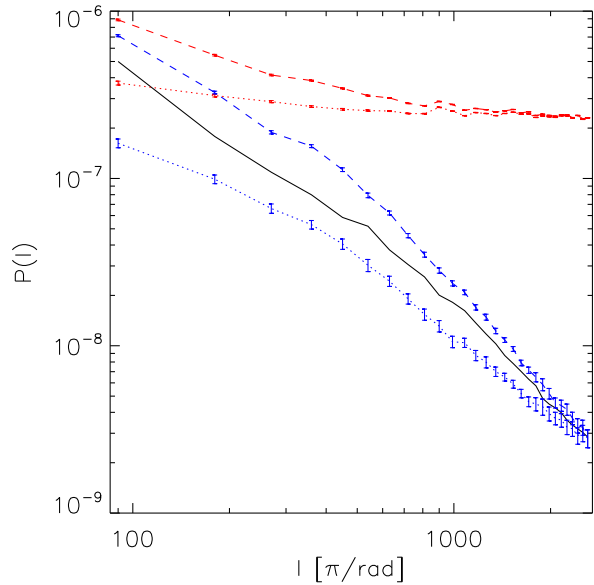}
  \caption{Power spectra (dashed) and error power spectra (dotted) for the reconstructions. The true convergence power spectrum (black solid line) is plotted for comparison. Red: maximum-likelihood approach. Blue: maximum-probability reconstruction including the phase prior. The reconstructions including the phase prior have $S/N>1$ even beyond $l=1000$, far into the domain where the shear data is noise-dominated.}
 \label{res_plot}
\end{figure}

\begin{figure}
  \includegraphics[width=0.5\textwidth]{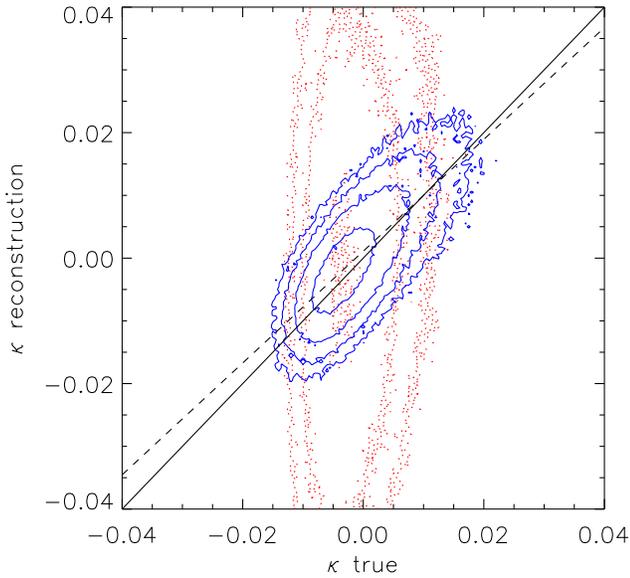}
  \caption{The contours for a 2D histogram of pixels in the reconstruction vs. pixels in the true convergence. Contours are for $10$,$10^{1.5}$,$10^{2}$,$10^{2.5}$ values, and the histogram shows a concatenation of reconstructions for 100 different fields. Results are shown with phase prior (blue solid) and maximum-likelihood approaches (red dotted). The best fit line to the phase reconstruction contours (black dashed) has a gradient of $0.89$ and offset of $0.001$.}
 \label{res_plot2}
\end{figure}

\begin{figure}
  \includegraphics[width=0.5\textwidth]{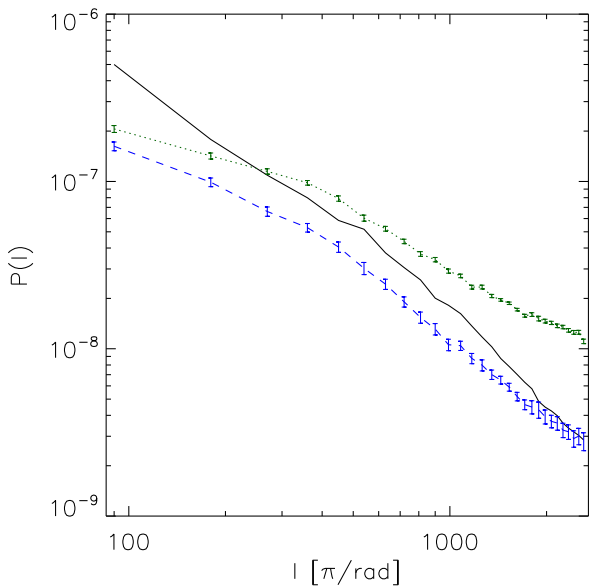}
  \caption{Dependence on the starting position. The true convergence power spectrum (black solid line) is plotted for comparison. We show the error power for a reconstruction starting from a $\kappa^{\rm{initial}}$ filtered according to Equation \ref{filter} (blue dashed) and an unfiltered one (green dotted).}
\label{init_plot}
\end{figure}

\begin{figure}
\includegraphics[width=0.5\textwidth]{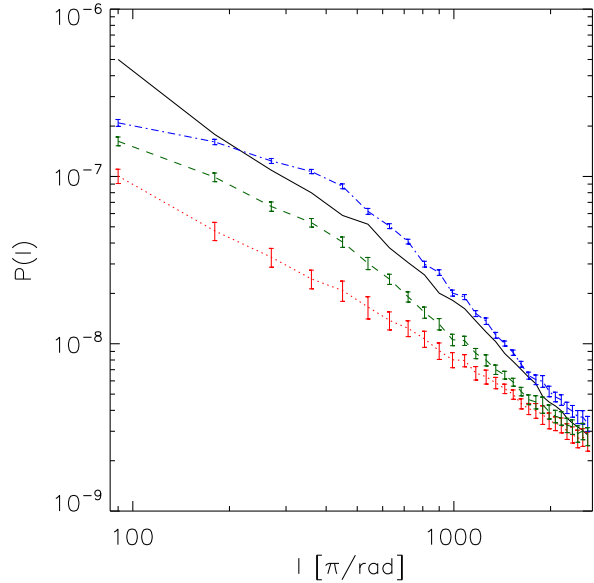}
\caption{Dependence on the phase tolerance. The true convergence power spectrum (black solid line) is plotted for comparison. The lines show errors for reconstructions with phase tolerance of $\sigma_{\alpha}=1.1\cdot{\rm{MAD}}_{\Delta\alpha}$ (red dotted), $\sigma_{\alpha}=2.2\cdot{\rm{MAD}}_{\Delta\alpha}$ (green dashed) and $\sigma_{\alpha}=3.3\cdot{\rm{MAD}}_{\Delta\alpha}$ (blue dot-dashed).}
 \label{tol_plot}
\end{figure}

\begin{figure}
  \centering
   \includegraphics[width=0.5\textwidth]{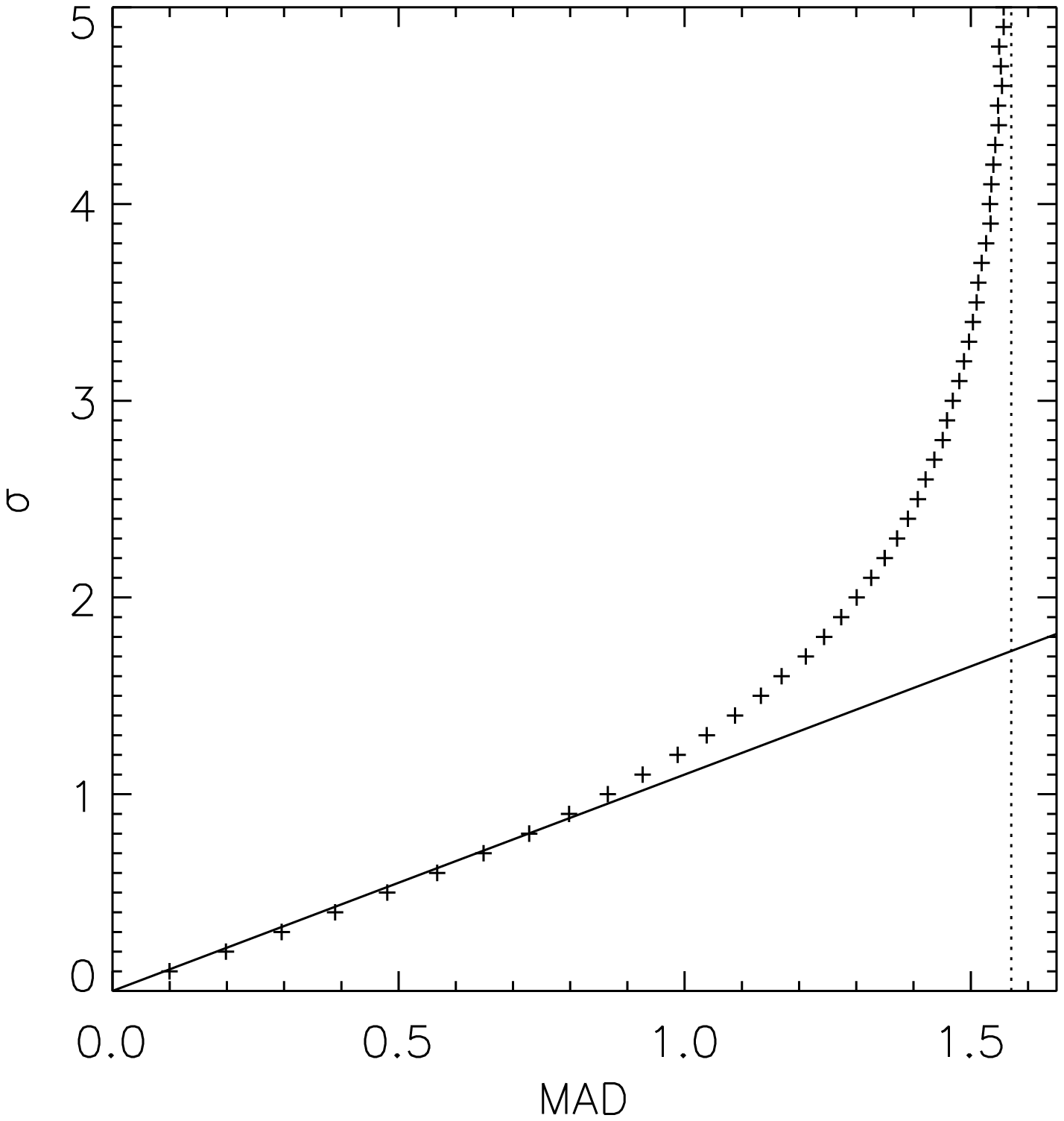}
  \caption{Median absolute deviation (${\rm{MAD}}$) as an estimator of $\sigma_\alpha$. Assuming $\sigma_{\alpha}$ can be estimated as $1.1\cdot{\rm{MAD}}_{\Delta\alpha}$ (solid line) is justified for values of $\sigma_{\alpha}\lesssim1$ (see Figure~\ref{ph_error}). For larger values, ${\rm{MAD}}_{\Delta\alpha}$ will tend to a~constant (here $\pi/2$).} 
  \label{mad}
\end{figure}

Figure~\ref{res_plot} shows the power spectra (dashed) and error power spectra (dotted) of the reconstruction averaged over $100$ fields. The maximum-likelihood reconstruction (red) is dominated by noise on most scales. Including the phase prior (blue) leads to a reconstruction that has higher $S/N$ than the ML reconstruction on all scales, and has $S/N>1$ even beyond $l=1000$, far into the domain where the initial shear data is noise-dominated. On a pixel by pixel basis the phase prior improves the correlation between the true convergence and the reconstruction as shown in Figure~\ref{res_plot2}. The Pearson correlation coefficient changes from $0.22$ for the ML reconstruction to $0.72$ in the case of the MP reconstruction.

The reduction of the noise visible in Figure 6 is due to the interplay between the galaxy phases and both the phase and amplitude of the lensing. Given noisy shear data, and if the phases of the two fields disagree strongly, the only permitted hypothesis that satisfies both the phase prior and the likelihood with modest probability, has low amplitude for the signal. On the other hand, where the phases agree, a higher amplitude is permitted.

To assess the errors on curves in Figure~\ref{res_plot}, an additional $100$ runs different starting points were performed on a single $2^{\circ} \times 2^{\circ}$ field, to see the variation in reconstructions permitted by the optimiser. The different $\kappa_i^{\rm{initial}}$ fields were generated by multiplying each mode in $\tilde\kappa_{\rm{g}}$ by a complex random number with each component drawn from a standard normal distribution $\mathcal{N}(0,1)$. The error bars on different power spectra in Figures~\ref{res_plot},~\ref{init_plot} and~\ref{tol_plot} show the standard deviation in error powers of this set of runs. We see that these errors are substantially smaller than the variation between the maximum-likelihood and maximum-probability runs (Figure~\ref{res_plot}) and also between maximum-probability runs with different values of the $\sigma_{\alpha}$ parameter (Figure~\ref{tol_plot}).

To check the dependence of the reconstruction on the initial guess $\kappa^{\rm{initial}}$, further reconstructions with the phase prior were performed. The phase tolerance was again set to $\sigma_\alpha=2.2\cdot{\rm{MAD}}_{\Delta\alpha}$ but $\kappa^{\rm{initial}}$ was left unfiltered. Figure~\ref{init_plot} shows the errors on these reconstruction compared to the analogous filtered one. The reconstruction with an unfiltered starting guess (green dotted) deviates more from the reconstruction with a filtered one (blue dashed) on small scales, $l\gtrsim 1000$ suggesting that the posterior probability surface is very flat in some directions (or multimodal). Although, the difference is visible on all scales, the reconstruction remains a substantial improvement over the maximum likelihood reconstruction in Figure \ref{res_plot}.

The tolerance we permit on the phases has a moderate impact on the reconstruction, as shown in Figure~\ref{tol_plot}. The lines show error power spectra for reconstruction with phase tolerance of $\sigma_{\alpha}=1.1\cdot{\rm{MAD}}_{\Delta\alpha}$ (red dotted), $\sigma_{\alpha}=2.2\cdot{\rm{MAD}}_{\Delta\alpha}$ (green dashed) and $\sigma_{\alpha}=3.3\cdot{\rm{MAD}}_{\Delta\alpha}$ (blue dot-dashed), and the error power grows by a factor of  two on intermediate scales between the tightest and weakest of these tolerances. However, independent of the phase tolerance the reconstructions are similar on small scales where the reconstruction is noise dominated, and on the largest scales where the likelihood term is large.

\section{Conclusions}
In this paper, we have proposed a maximum-probability reconstruction method for the lensing convergence, and have studied the impact of a physically motivated prior term. 

To put a weakly informative prior on the Fourier phases of the modes, we made a prediction of the convergence from the galaxy number overdensity, and used this to inform the preferred phases of the reconstructed convergence field. In this way, by using only the phases of this field, we avoid the use of the unknown amplitude of the linear galaxy bias. We also do not require a deterministic bias, as we allow a phase deviation between the galaxy distribution and the underlying matter density.

By implementing and testing this method with a realistic simulated galaxy shear catalogue, we have found that a weak prior on phases provides a good quality 2-D density reconstruction with signal-to-noise $S/N \geq 1$ on scales up to and beyond $l=1000$ (Figure~\ref{res_plot}). 

The sensitivity of the phase prior reconstruction to initial conditions (Figure~\ref{init_plot}) shows that the probability surface is flat in directions associated with noise dominated modes, as expected. However, an approximate knowledge of the power spectrum can help to select a solution with modest signal-to-noise even on the smallest scales. The phase difference tolerance can be made more or less strict, depending on whether one wishes to make a joint reconstruction using weak lensing and phases, or instead to make a reconstruction from weak lensing weakly informed by phases. In either case, the reconstruction is found to be an improvement over maximum likelihood reconstruction (contrast Figures \ref{tol_plot} and \ref{res_plot}). 

Although, most of the phase information is coming from the galaxy field, the amplitude of the modes is determined by the interplay between these and the lensing, which includes both phase and amplitude information. It is important to emphasise that in Figure~\ref{maps}(d) the amplitude is an assumption, whereas in Figure~\ref{maps}(c) it is derived purely from data.

In summary, using the phase information from the galaxy distribution to inform weak lensing density reconstruction, appears to be a very powerful addition to the tools we can use for mass mapping. As these maps combine information from the weak lensing and galaxy fields, they can potentially be used to improve our understanding of the relation between dark matter and galaxies, i.e. the bias.

\section*{Acknowledgments}
We thank Bruce Bassett, Mathew Becker, Rob Crittenden, Alan Heavens and Phil Marshall for useful discussions. RS also thanks B.~ Bassett for organising the Cape Town Cosmology School 2012 which inspired some of the ideas presented in this paper. 

This work was partially supported by STFC grant ST/K00090X/1 and a Royal Society-NRF International Exchange Grant.  RS acknowledges support from STFC in the form of a Research Studentship. PM is supported by the U.S. Department of Energy under Contract No. DE- FG02-91ER40690.
\\\\
Please contact the authors to request access to research materials discussed in this paper.
\bibliography{lib}
\bibliographystyle{mn2e}

\appendix

\section{Wrapped Cauchy distribution}
\label{wr-c}
The Cauchy pdf is given by
\begin{equation}
f_{\rm{C}}(x;x_0,\sigma)=\frac{1}{\pi}\cdot\frac{\sigma}{\sigma^2+(x-x_0)^2},\;x\subset(-\infty,\infty).
\end{equation}
The Wrapped Cauchy pdf is defined as
\begin{equation}
f_{\rm{WC}}(\beta;\beta_0,\gamma)=\sum^\infty_{n=-\infty}\frac{\sigma}{\pi(\sigma^2+(x-x_0+2\pi n)^2)},
\end{equation}
which gives
\begin{equation}
f_{\rm{WC}}(\beta;\beta_0,\gamma)=\frac{1}{2\pi}\cdot\frac{1-\gamma^2}{1+\gamma^2+2\gamma\cos(\beta-\beta_0)},
\end{equation}
where $\gamma=e^{-\sigma}$ and $\beta\subset[-\pi,\pi)$.

A Cauchy distributed random number $x$ can be generated from two independent normally distributed random numbers $y_1,y_2\sim\mathcal{N}(0,1)$ as 
\begin{equation}
x=x_0+\sigma\frac{y_1}{y_2}.
\end{equation}
Then a Wrapped Cauchy distributed random number is obtained by taking
\begin{equation}
\beta=x\;{\rm{mod}}\;2\pi,
\end{equation}
and applying a procedure similar to the one in Section~\ref{sec:phdist}, i.e., if $\beta$ is less than $-\pi$, we add $2\pi$ to $\beta$; if $\beta$ is greater than or equal to $\pi$ then we subtract $2\pi$ from $\beta$.

For a distribution with $\beta_0=0$ the parameter $\sigma_{\alpha}$ can be approximated (for small values) as $1.1\cdot{\rm{MAD}}_{\Delta\alpha}$ (Figure~\ref{mad}). For high values of $\sigma_{\alpha}$ this approximation breaks down; as $\sigma_{\alpha}\rightarrow\infty$ the Wrapped Cauchy tends to a uniform distribution, and ${\rm{MAD}}_{\Delta\alpha}$ goes to a constant equal to the standard deviation of the uniform distribution, here $\pi/2$ (see Figure~\ref{mad}).

\bsp

\label{lastpage}

\end{document}